\documentclass[12pt]{iopart}
\usepackage{graphicx}
\usepackage{epsfig}
\usepackage{subfigure}
\usepackage[small,bf]{caption}
\begin{document}

\title{Bulk Properties of Pb-Pb collisions at $\sqrt s_{NN}=2.76$~TeV measured by ALICE}
\author{Alberica Toia for the ALICE Collaboration}

\address{CERN Div. PH, 1211 Geneva 23}
\ead{alberica.toia@cern.ch}
\begin{abstract}
  Global variables, such as the charged particle multiplicity and the
  transverse energy are important observables to characterize
  Relativistic Heavy Ion collisions and to constrain model
  calculations. The charged particle multiplicity $dN_{ch}/d\eta$ and
  transverse energy $dE_T/d\eta$ are measured at $\sqrt
  s_{NN}=2.76$~TeV in Pb-Pb collisions as a function of centrality and
  in pp collisions. The fraction of inelastic cross section seen by
  the ALICE detector is calculated either using a Glauber model or the
  data corrected by simulations of nuclear and electromagnetic
  processes, or data collected with a minimum bias interaction
  trigger. The centrality, defined by the number of nucleons
  participating in the collision, is obtained, via the Glauber model,
  by relating the multiplicity distributions of various detectors in
  the ALICE Central Barrel and their correlation with the spectator
  energy measured by the Zero-Degree Calorimeters.  The results are
  compared to corresponding results obtained at the significantly
  lower energies of the BNL AGS, the CERN SPS, and the BNL RHIC, and
  with models based on different mechanisms for particle production in
  nuclear collisions. Particular emphasis will be given to a
  discussion on systematic studies of the dependence of the centrality
  determination on the details of the Glauber model, and the validity
  of the Glauber model at unprecedented collision energies.
\end{abstract}

\section{Introduction: the importance of bulk properties}
One fundamental element to study ultrarelativistic collisions is the
characterization of the interaction in terms of bulk variables such as
the transverse energy and the number of charged particles.  These
variables are closely related to the collision geometry and are
important in understanding global properties of the system during the
collision.  This paper presents the first study of $dN_{ch}/d\eta$ at
mid- and forward-rapidity and $dE_T/d\eta$ measured at
$\sqrt s_{NN}=2.76$~TeV by the ALICE experiment. The centrality
dependence of these observables is characterized by the number of
participants N$_{part}$, determined with a Glauber model of Pb-Pb
collisions and is studied as a function of energy by comparing our
results to those at RHIC and at the SPS. Finally, the results are
compared to the available models.
Charged particles $N_{ch}$ and transverse energy $E_T$ are generated
by the initial scattering of the partonic constituents of the incoming
nuclei and possibly also by reinteractions among the produced partons
and hadrons. Without significant reinteraction, the observed
transverse energy is the same as that produced by the initial
scattering. The amount of the reinteraction, and therefore of the
equilibration of the fireball, can be measured evaluating the decrease
of the initial scattering energy. To some extent transverse
hydrodynamic flow can compensate this equilibrium.  However gluon
saturation can delay the onset of the flow reducing the effective
pressure and thereby reducing the difference in the initially produced
and the observed energy.  The systematic study of global event
properties, in particular transverse energy, charged multiplicity and
mean transverse momentum, their centrality and energy dependence, may
impose significant constraints on the collision dynamics.

\section{Measurement of the centrality}
ALICE has collected in the Pb-Pb run last year $\sim$30 million events
with nuclear collisions using minimum bias triggers with increasingly
tigther conditions.  These triggers use the VZERO scintillators
(covering $2.8<\eta<5.1$ and $-3.7<\eta<-1.7$) and Silicon Pixel
Detector (SPD, $|\eta|<1.4$)\cite{alice-det}, and their efficiency
ranges from 97 to 99\%, as measured in simulations and dedicated pp
runs taken with minimum interaction triggers. 
However peripheral collisions are strongly contaminated by electromagnetic
(EM) background for which the cross sections are much higher than at
the lower energies measured so far.  QED processes and photo-nuclear
interactions, where one photon from the EM field of one of the nuclei
interacts with the other nucleus, possibly fluctuating to a vector
meson yield soft particles at mid-rapidity \cite{Djuvsland:2010qs}. At beam
rapidity the cross section is large for electromagnetic dissociation,
where one or both nuclei breaks up as a consequence of the EM
interaction.

\subsection{Glauber Model}
\begin{figure}[btp]
  \includegraphics[width=0.5\textwidth]{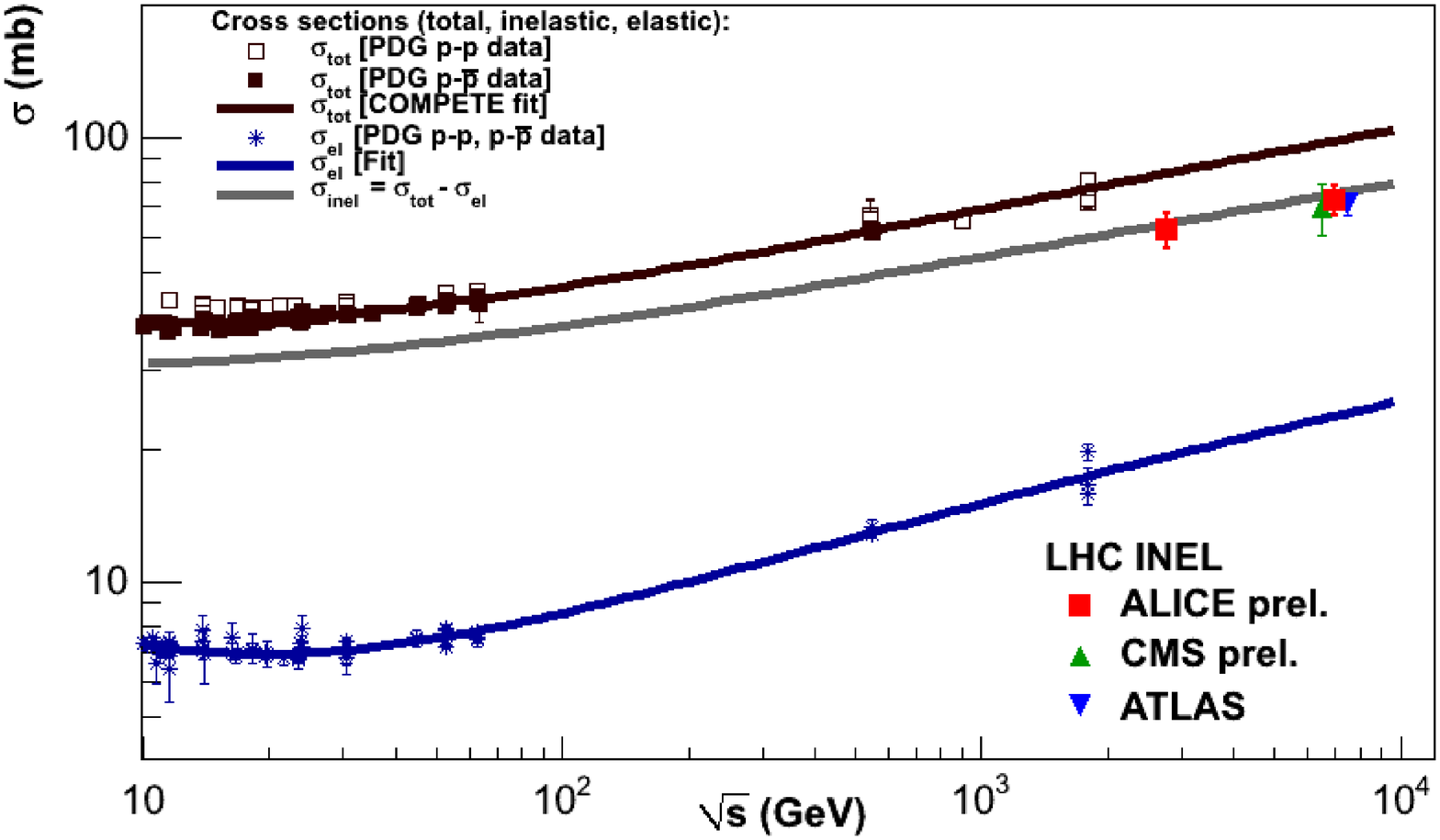}
  \includegraphics[width=0.5\textwidth]{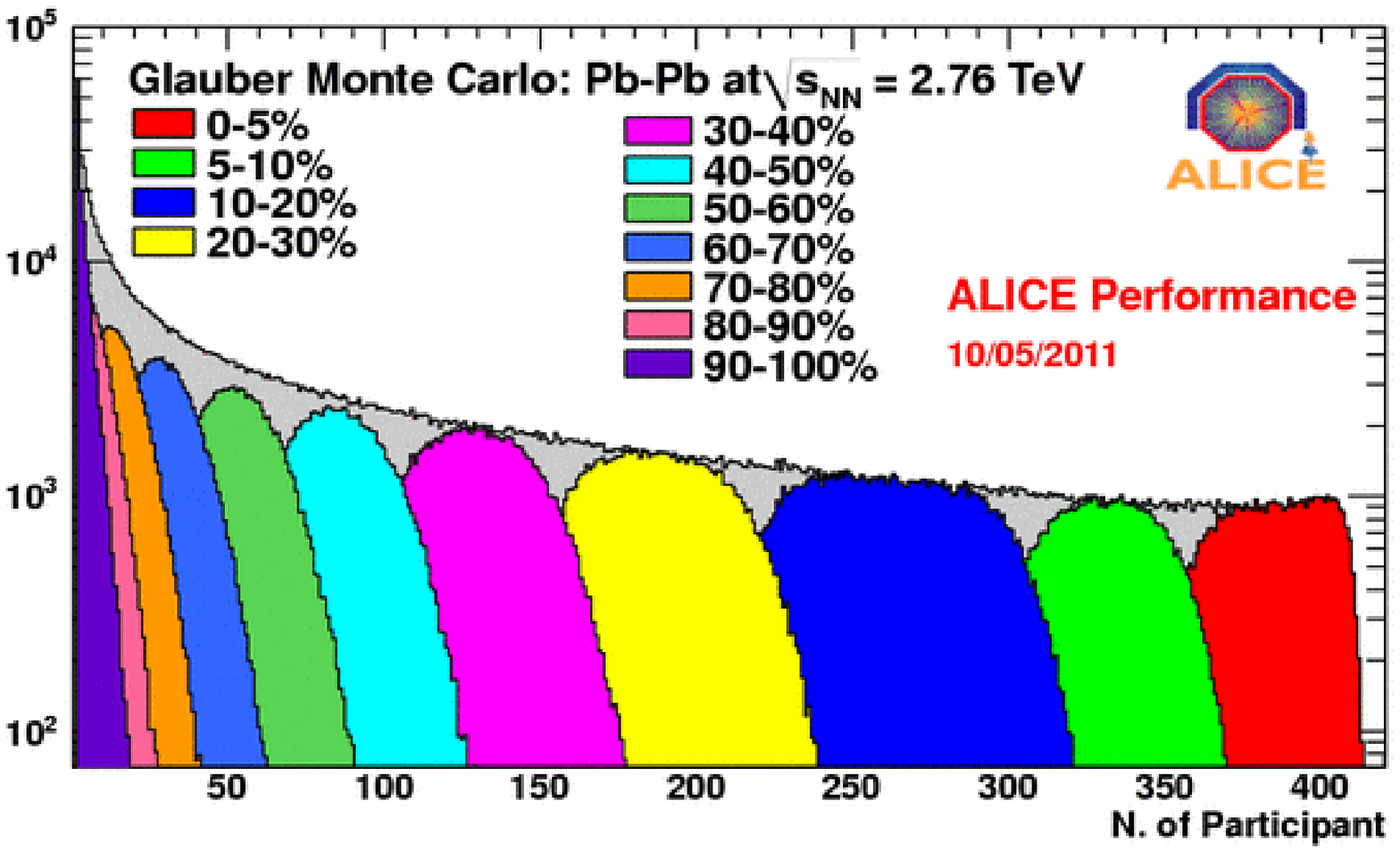}
  \caption{(left) World-data of total (squares) and elastic (star) cross sections in $pp$ and $p\bar{p}$ collisions. The preliminary data from ALICE, ATLAS and CMS agree well with the extrapolation for the $\sigma_{inel}$. (right) Number of participants N$_{part}$ for the centrality classes used in the analysis obtained for percentiles of the geometric cross section.
  \label{fig:glau}}
\end{figure}
Hadronic processes are described in a simple geometrical picture by
the Glauber Model, which assumes straight−line nucleon trajectories
and N-N cross section independent of the number of collisions the
nucleons have undergone before. The nuclear density profile is given
by a Woods-Saxon distribution.  We used a nucleon-nucleon cross
section, see Fig.~\ref{fig:glau} (left) of 64~mb, estimated from
all available measurements (before LHC) \cite{den11} and now confirmed
by measurements by ATLAS, CMS and ALICE ones, thanks to Van der Meer
scans \cite{oya11} and a detailed analysis of the diffractive systems
\cite{pog11}.  The collision geometry determines the number of
nucleons that participate in the reaction, so-called number of
participants N$_{part}$, in Fig.~\ref{fig:glau} (right). 
The definition of centrality is based on the basic assumption that the impact
parameter $b$ is monotonically related to particle multiplicity or
the energy produced at mid-rapidity.

\subsection{Multiplicity distributions in the Central Barrel}
\begin{figure}[ht]
  \includegraphics[width=0.45\textwidth]{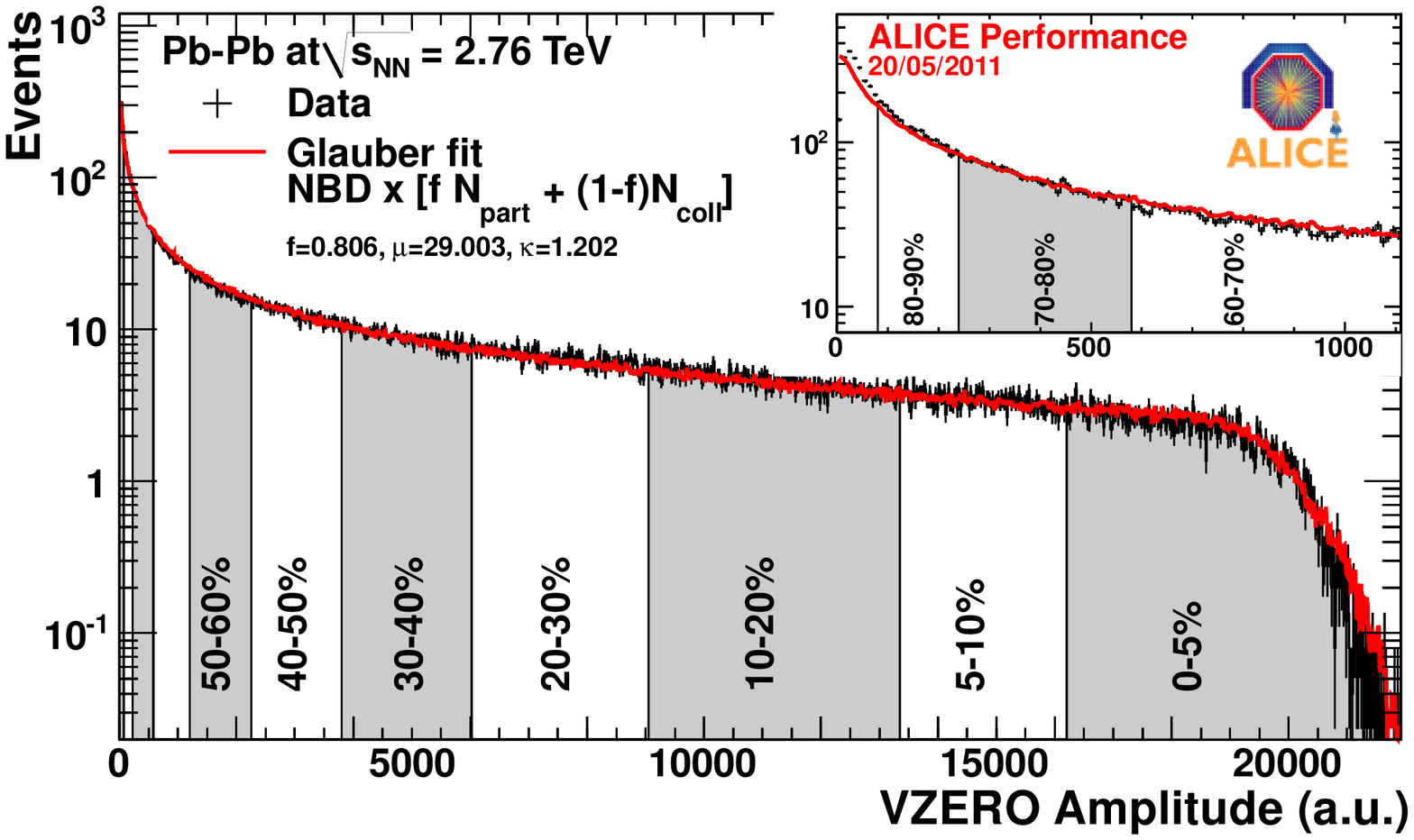}
  \includegraphics[width=0.42\textwidth]{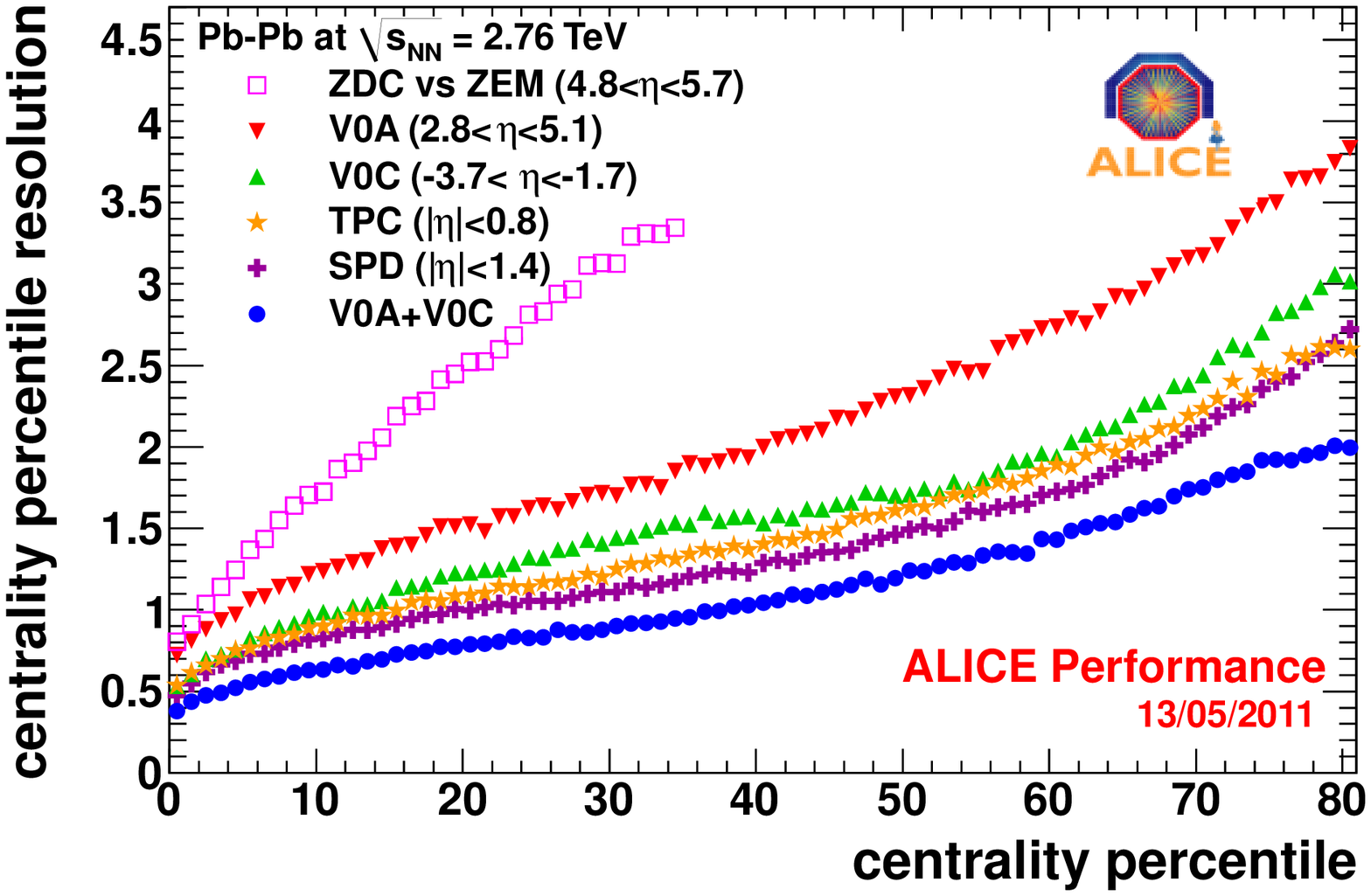}
  \caption{(left) Distribution of the sum of amplitudes in the VZERO scintillators. The line shows the fit of the Glauber calculation to the measurement. The centrality classes used in the analysis are indicated in the figure. The inset shows a zoom of the most peripheral region. (right) Centrality resolution for all the estimators evaluated in the analysis.  }
  \label{fig:vzeroglau}
\end{figure}

The multiplicity distribution of any of these centrality observables
has a classical shape. As an example, the distribution of VZERO
amplitude is shown in Fig.~\ref{fig:vzeroglau} (left). The peak
corresponds to most peripheral collisions (contaminated by EM
background and by missing events due to the trigger efficiency), the
plateau to the mid-central and the edge to the central collisions
which results from the intrinsic fluctuations and the detector
acceptance and resolution.  Other observables include the number of
Time Projection Chamber (TPC, $|\eta|<0.9$) tracks and the number of
SPD clusters.  The VZERO distribution is fitted using a
phenomenological approach based on the Glauber Monte Carlo plus a
convolution of a model for the particle production and a negative
binomial distribution (NBD).  It is assumed that the number of
independently decaying precursor particles (``ancestors'') is given by
a 2 component model $N_{ancestors} = f \cdot N_{part} + (1-f) \cdot
N_{coll}$, where $f\sim 80$~\% from the fit quantifies their relative
contribution. Other ancestor dependences have been tested, using
power-law functions of N$_{part}$ or N$_{coll}$.  The number of
particles produced per precursor source was assumed to follow a NBD
distribution. The fit is performed in a region corresponding to ~88\%
of the total cross section to avoid the peak of contamination and
inefficiency.  Extracting the number of participants from the Glauber
fit gives accesses directly to the N$_{part}$, nearly identical to the
geometrical one. However it is important to remember that we use the
Glauber model and the ancestor assumptions only to determine the
fraction of total cross section that we see, confirming the results
obtained with the data-based analysis.

\subsection{Spectator measurement with the Zero Degree Calorimeter (ZDC)}
Another way to determine the centrality is to measure the energy
deposited by the spectators in the Zero Degree Calorimeter (ZDC). This
in principle provides directly the number of participants, however the
nuclear fragmentation breaks the simple relation in the measured
variables.  The ZDC therefore needs to be correlated to another
detector, in this case the electromagnetic calorimeter ZEM. Since the
ZDC is far from the interaction point and therefore rather independent
on the vertex, this centrality measure is particularly suited for the
analysis that does not require any vertex cut, and it gives good
results for central collisions, where the ZDC signal is well
correlated with the number of spectators.

\subsection{Centrality Resolution}
We evaluated the performance of the centrality determination by
comparing all our estimates. This is shown in
Fig.~\ref{fig:vzeroglau} (right). The figure shows the RMS of the (Gaussian)
distribution of $\sigma_i = cent_i - <cent>$, where $<cent>$ is
calculated iteratively by
\begin{equation}
<cent> = \frac{\sum cent_i / \sigma_i^2(cent_i)}{\sum 1 / \sigma_i^2(cent_i)}
\end{equation}
The resolution depends on the rapidity coverage of the
detector used. So when scaled by the square root of the $N_{ch}$ measured
in that detector all the results line up together. The resolution ranges from 0.5\% in central to 2\% in peripheral collisions.

\section{Measurement of the charged particle multiplicity at mid-rapidity}
\begin{figure}[tp]
  \begin{minipage}[b]{1.0\linewidth}
  \includegraphics[width=0.45\textwidth]{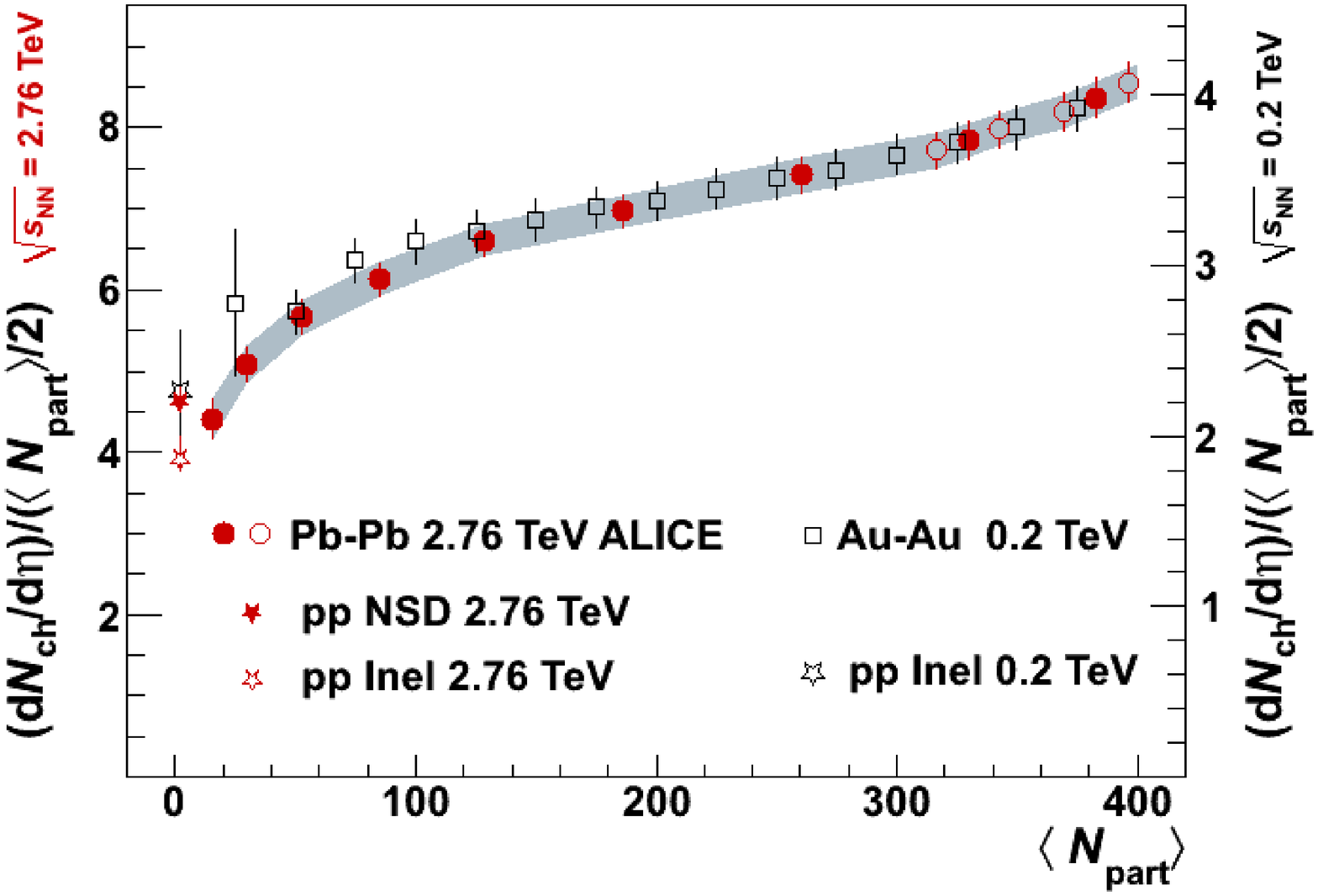}
  \includegraphics[width=0.45\textwidth]{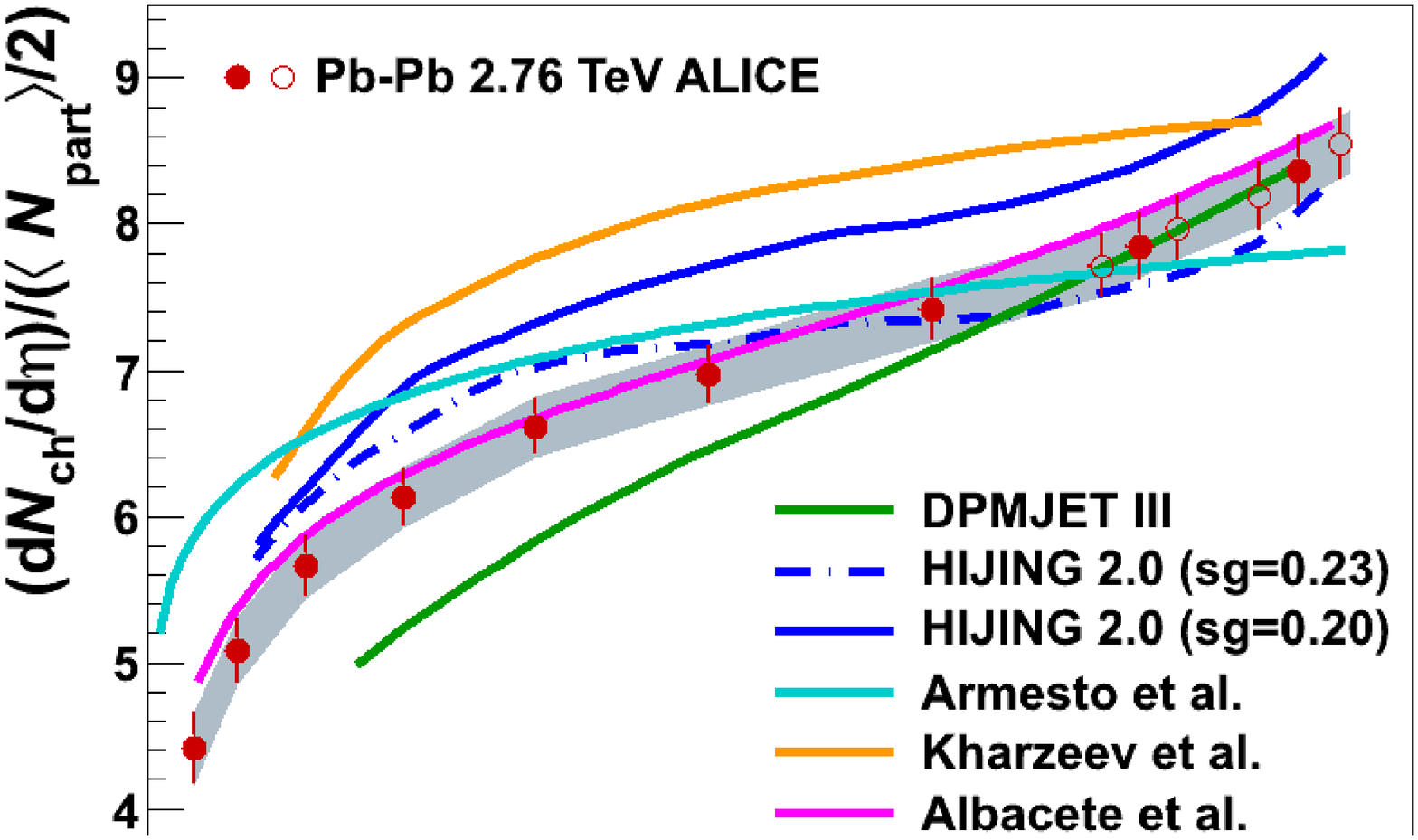}
  \end{minipage} \hfill
  \begin{minipage}[b]{1.0\linewidth}
    \caption{(left) Charged particle pseudo-rapidity density $(dN_{ch}/d\eta)/(<N_{part}>/2)$ per participant pair as function of N$_{part}$ for Pb-Pb collisions at 2.76~TeV and Au-Au collisions at 200~GeV (RHIC average) \cite{mulRHIC}. The scales for the different data sets are shown on the left and on the right respectively. The values for the pp collisions (LHC) are interpolations between data at 2.36 and 7~TeV. The pp value at RHIC energy is from \cite{rhicpp}. (right) $(dN_{ch}/d\eta)/(<N_{part}>/2)$ compared with model calculations (see Sec.~\ref{sec:discuss}).
  \label{fig:dNch}}
  \end{minipage} 
\end{figure}

The multiplicity of charged particles is measured at mid rapidity with
an analysis based on SPD tracklets \cite{mul11,loi11}. Figure~\ref{fig:dNch} (left) shows
the charged particle pseudorapidity density per participant pair in
$|\eta|<$1 as a function of N$_{part}$.  The Pb-Pb data points
extrapolate well to the pp measurement.  Compared to the RHIC results \cite{mulRHIC},
also shown in the figure, there is an increase of about a factor 2.1,
but the centrality trend looks very similar, at least for
N$_{part}>100$. Also the RHIC points match well the corresponding pp
measurement \cite{rhicpp}. One can observe a splitting in the centrality
dependence for N$_{part}<100$.

\section{Estimate of the transverse energy}
\begin{figure}
    \begin{minipage}[b]{0.7\linewidth}
      \includegraphics[width=0.45\textwidth]{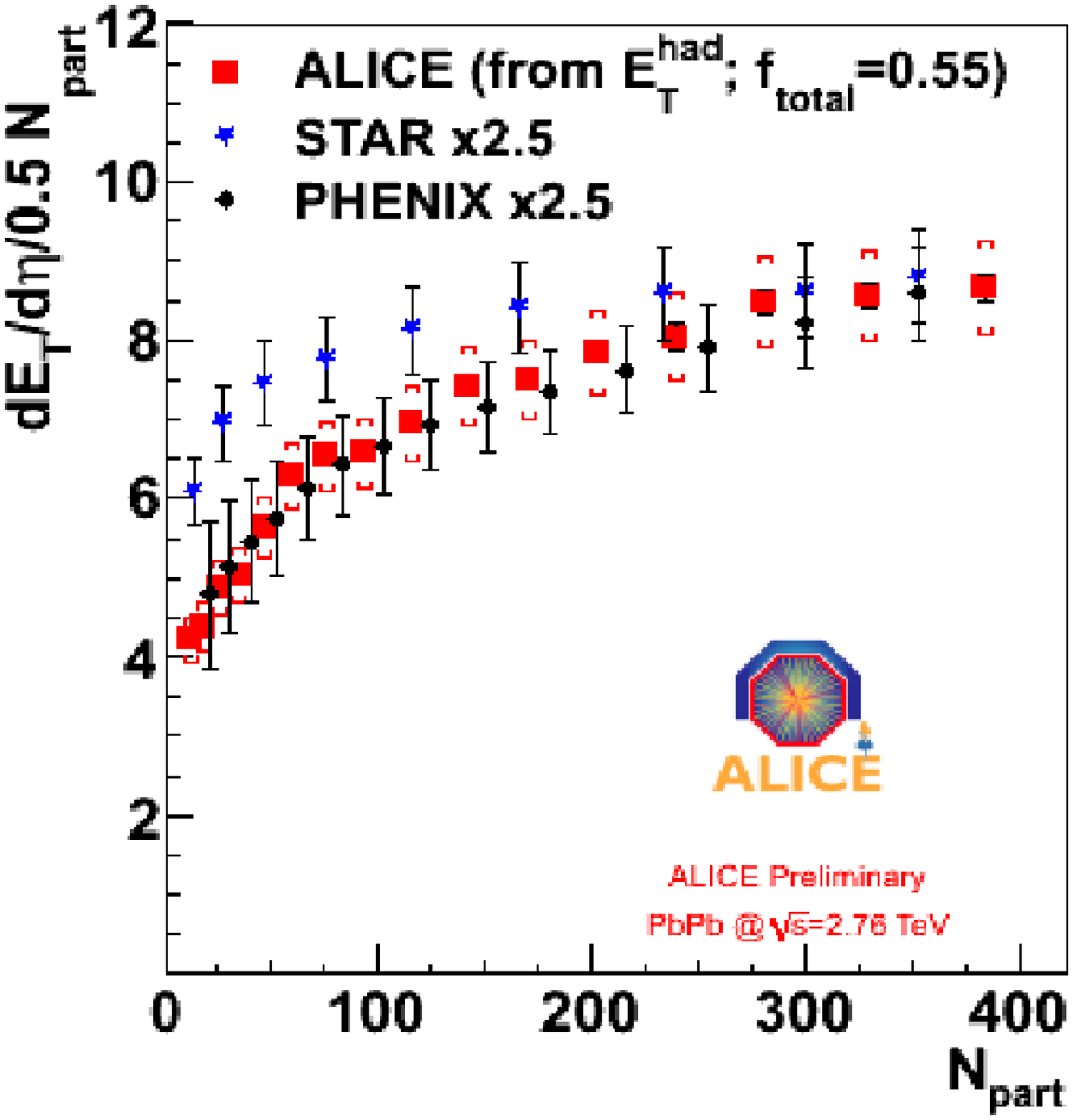}
      \includegraphics[width=0.43\textwidth]{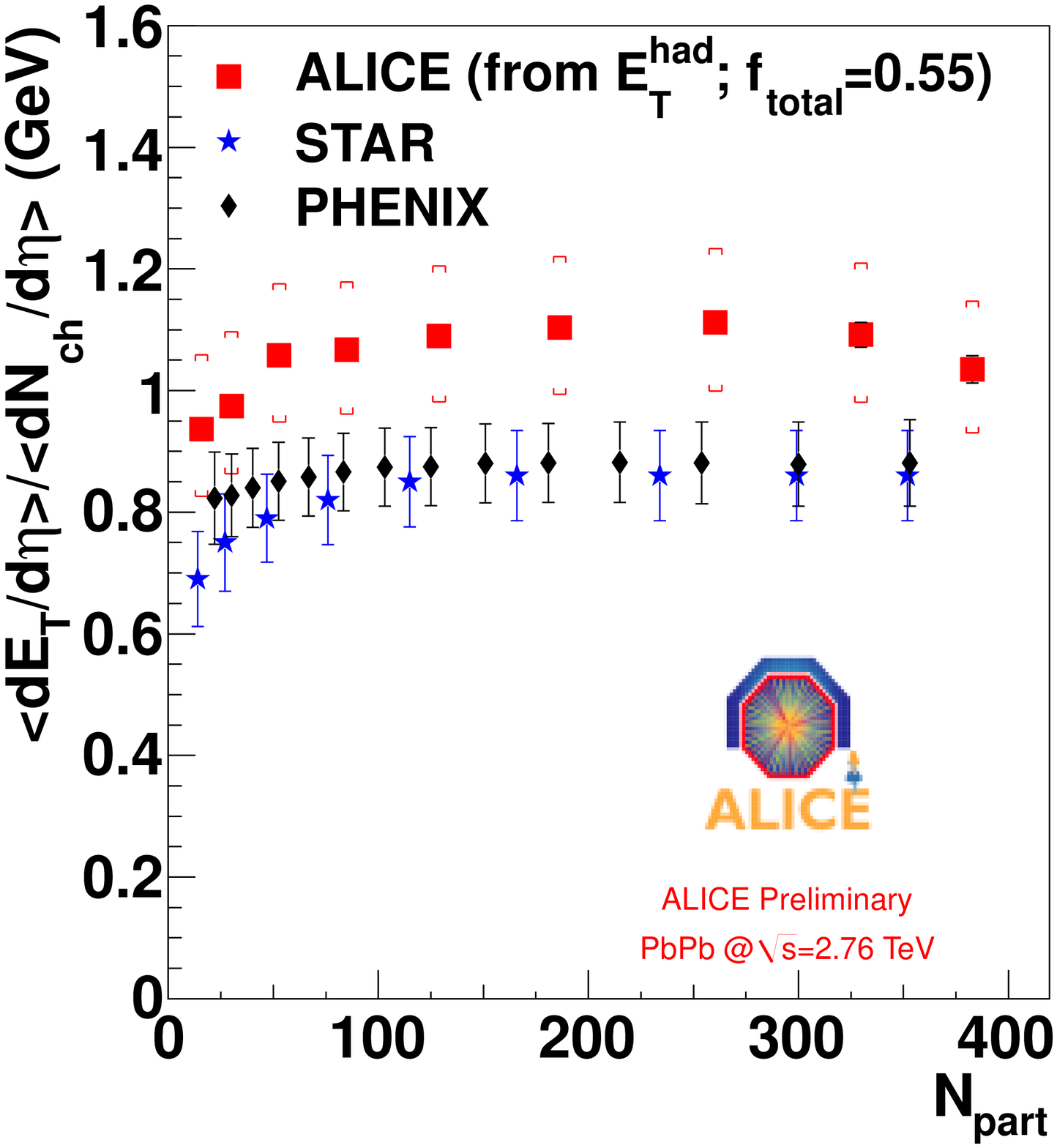}
    \end{minipage}
\hspace{-1cm}
    \begin{minipage}[b]{0.35\linewidth}
      \caption{(left) Transverse energy per participant pairs $(dE_{T}/d\eta)/(<N_{part}>/2)$ as function of N$_{part}$ for Pb-Pb collisions at 2.76~TeV and Au-Au collisions at 200~GeV. The RHIC data are multiplied by a factor of 2.5 \cite{mulRHIC}. (right) Ratio of $(dE_{T}/d\eta)$ and $(dN_{ch}/d\eta)$ as function of N$_{part}$. 
      \label{fig:dET}}
    \end{minipage}
\end{figure}

The transverse energy is estimated by measuring the charged hadronic
energy with the tracking system corrected by the fraction of neutral
particles not accessible by the tracking detectors.  The total energy
per participant pair is shown in Fig.~\ref{fig:dET} as a function of
N$_{part}$. As the multiplicity, it shows a steady rise with the
number of participants, very similar to that shown at RHIC but
increased by a factor of 2.5.  This measurement can be used to extract
the energy density using the Bjorken estimate
\begin{equation}
\epsilon=\frac{1}{\pi R^2 \tau}\frac{dE_T}{dy}
\end{equation}
where $\tau$ is the formation time and $\pi R^2$ is the effective area
of the collision.  The most central (0-5\%) value of $dE_T/dy$ gives
an $\epsilon\tau\sim 16$~GeV/(fm$^2c$) at LHC, about a factor 3 larger
than the corresponding one at RHIC \cite{mulRHIC}. Assuming that the
multiplicity is proportional to the entropy of the final state, and
that $\epsilon \propto T^4$, the factor 3 increase in energy density
corresponds to a 30\% increase in the temperature of the quark-gluon
plasma produced in these collisions (at the same time after the onset
of the reaction) compared with RHIC.

\section{Energy dependence of $(dN_{ch}/d\eta)$ and $(dE_{T}/d\eta)$}
\begin{figure}[tp]
  \begin{minipage}[b]{1.0\linewidth}
  \includegraphics[width=0.4\textwidth]{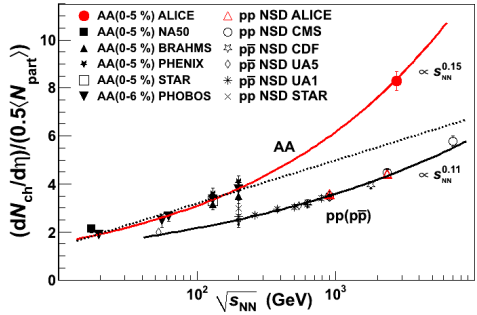}
  \hspace{1cm}
  \includegraphics[width=0.47\textwidth]{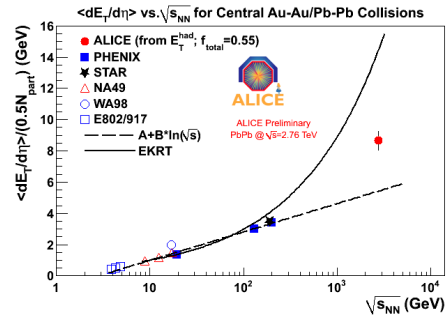}
  \end{minipage} \hfill
  \begin{minipage}[b]{1.0\linewidth}
    \begin{minipage}[b]{0.4\linewidth}
      \includegraphics[width=1.0\textwidth]{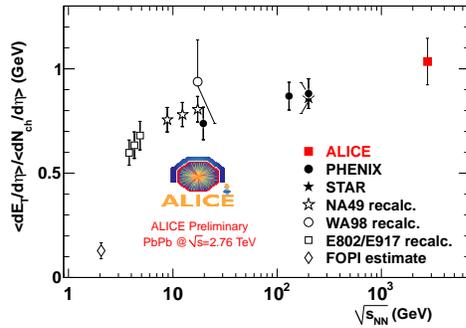}
    \end{minipage}
    \begin{minipage}[b]{0.6\linewidth}
      \caption{Top: Charged particle pseudorapidity density $(dN_{ch}/d\eta)/(<N_{part}>/2)$ (left) and transverse energy $(dE_{T}/d\eta)/(<N_{part}>/2)$ (right) per participant pairs for the most central central nucleus-nucleus and non-single diffractive pp collisions as a function of $\sqrt s_{NN}$. Bottom: Ratio of $(dE_{T}/d\eta)$ and $(dN_{ch}/d\eta)$ as a function of 
        $\sqrt s_{NN}$.
        \label{fig:sqrts}}
    \end{minipage}
  \end{minipage} 
\end{figure}

The energy dependence is analyzed by comparing the results in the most
central (0-5\%) centrality bin at LHC energy with those for the same
centrality bin at lower energy \cite{mulRHIC}. Note that in contrast
to RHIC data, our results are corrected for contamination by weak
decay products. Fig.~\ref{fig:sqrts} (top left panel) shows that the
charged particle multiplicity increases by a factor 2.1 with respect
to RHIC data, but only by 1.9 in pp collisions at similar energy. The
growth with energy is therefore different in pp and AA collisions,
confirming the interplay of N$_{part}$ and N$_{coll}$ dependence in
the particle production mechanism in heavy ion collisions.
\begin{figure}[tp]
  \begin{minipage}[b]{0.4\linewidth}
  \includegraphics[width=1.0\textwidth]{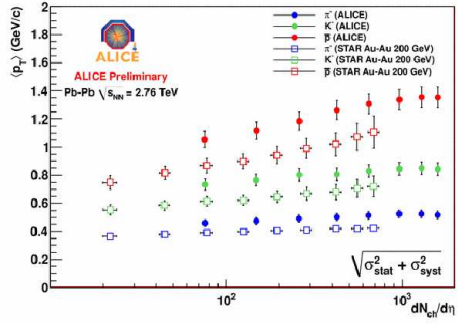}
  \end{minipage} 
  \begin{minipage}[b]{0.6\linewidth}
    \caption{Mean $p_T$ as a function of charged-particle density of identified hadrons ($\pi, K, p$) measured by ALICE at LHC energy of 2.76 TeV and by PHENIX and STAR at RHIC in Au-Au collisions at 200 GeV 
  \label{fig:meanmom}}
  \end{minipage} 
\end{figure}
The increase in transverse energy
(top right panel) is a factor 2.7 (consistent with the 2.5 reported on
Fig.~\ref{fig:dET} and a 5\% increase on the N$_{part}$ with respect
to RHIC), consistent with the observed increase of the mean particle
momentum (see Fig.~\ref{fig:meanmom}.

The energy dependency of both $(dE_{T}/d\eta)$
and $(dN_{ch}/d\eta)$ exhibits a power-law scaling (indicated by the red line in
Fig.~\ref{fig:sqrts} top left panel) stronger than the logarithmic scaling
(dotted lines in Fig.~\ref{fig:sqrts} top panels) suggested
by the lower energy experiments. The centrality evolution looks
similar for energy and multiplicity: both increase with energy and
with centrality.  Taking the ratio of $(dE_{T}/d\eta)$ to
$(dN_{ch}/d\eta)$ (see Fig.~\ref{fig:dET} (right panel)) one can
confirm the consistent behavior of the two observables as the ratio
$E_T/N_{ch}$ is rather independent on centrality, as was at RHIC, and
slightly increases with energy (see Fig.~\ref{fig:sqrts} (bottom
panel)).

\section{Measurement of the charged particle multiplicity at forward-rapidity}
\begin{figure}[btp]
  \includegraphics[width=0.33\textwidth]{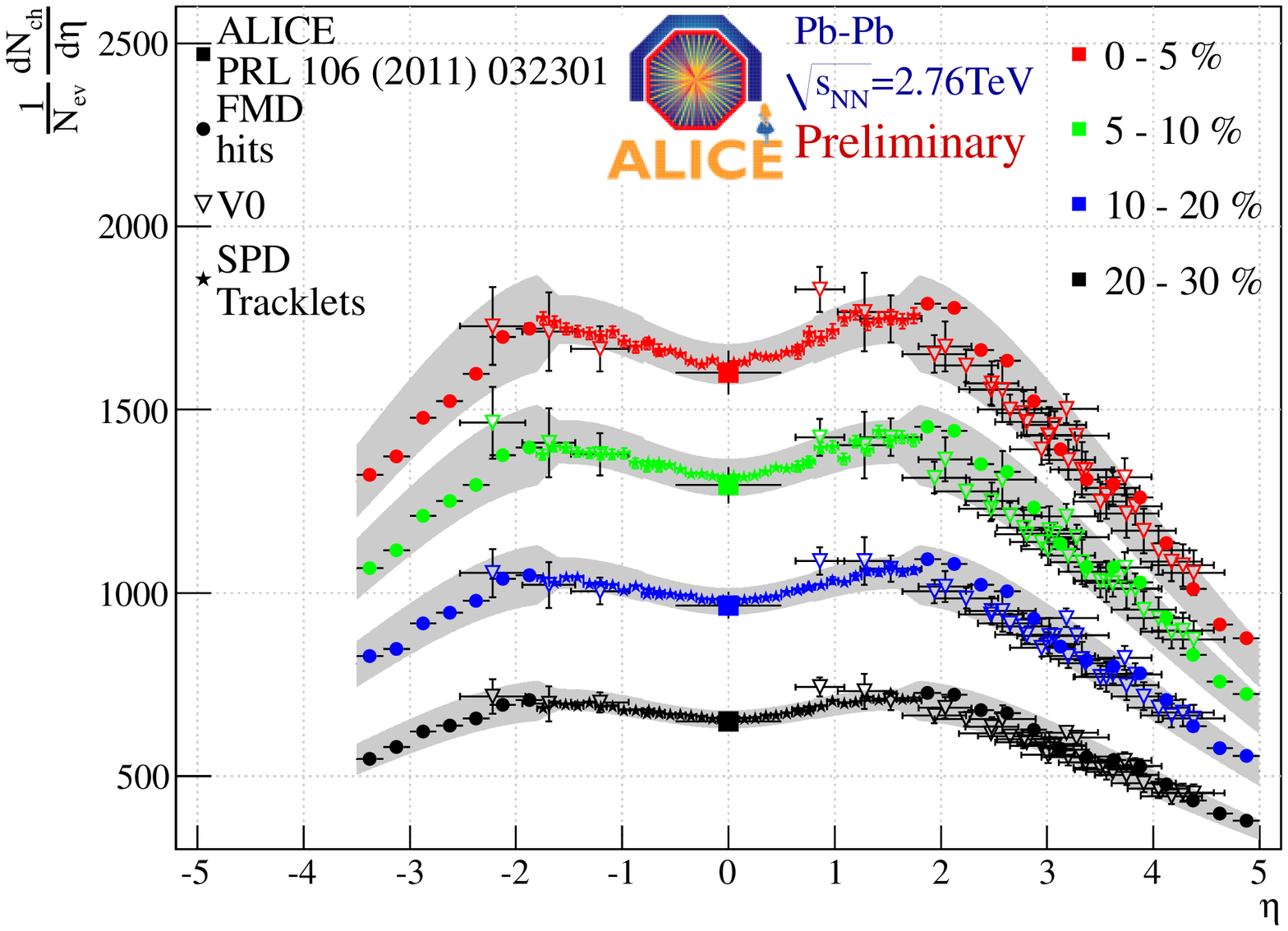}
  \includegraphics[width=0.33\textwidth]{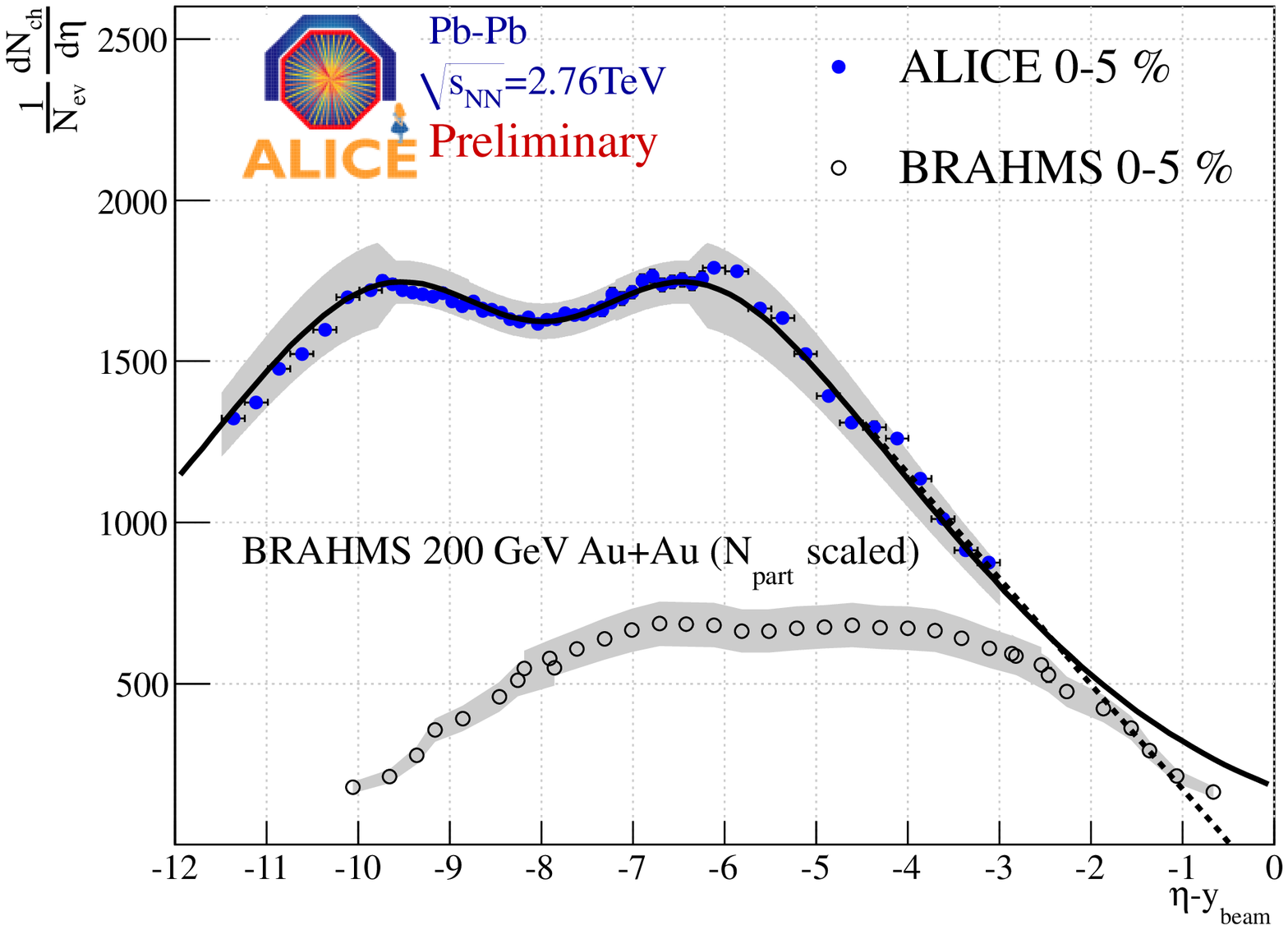}
  \includegraphics[width=0.32\textwidth]{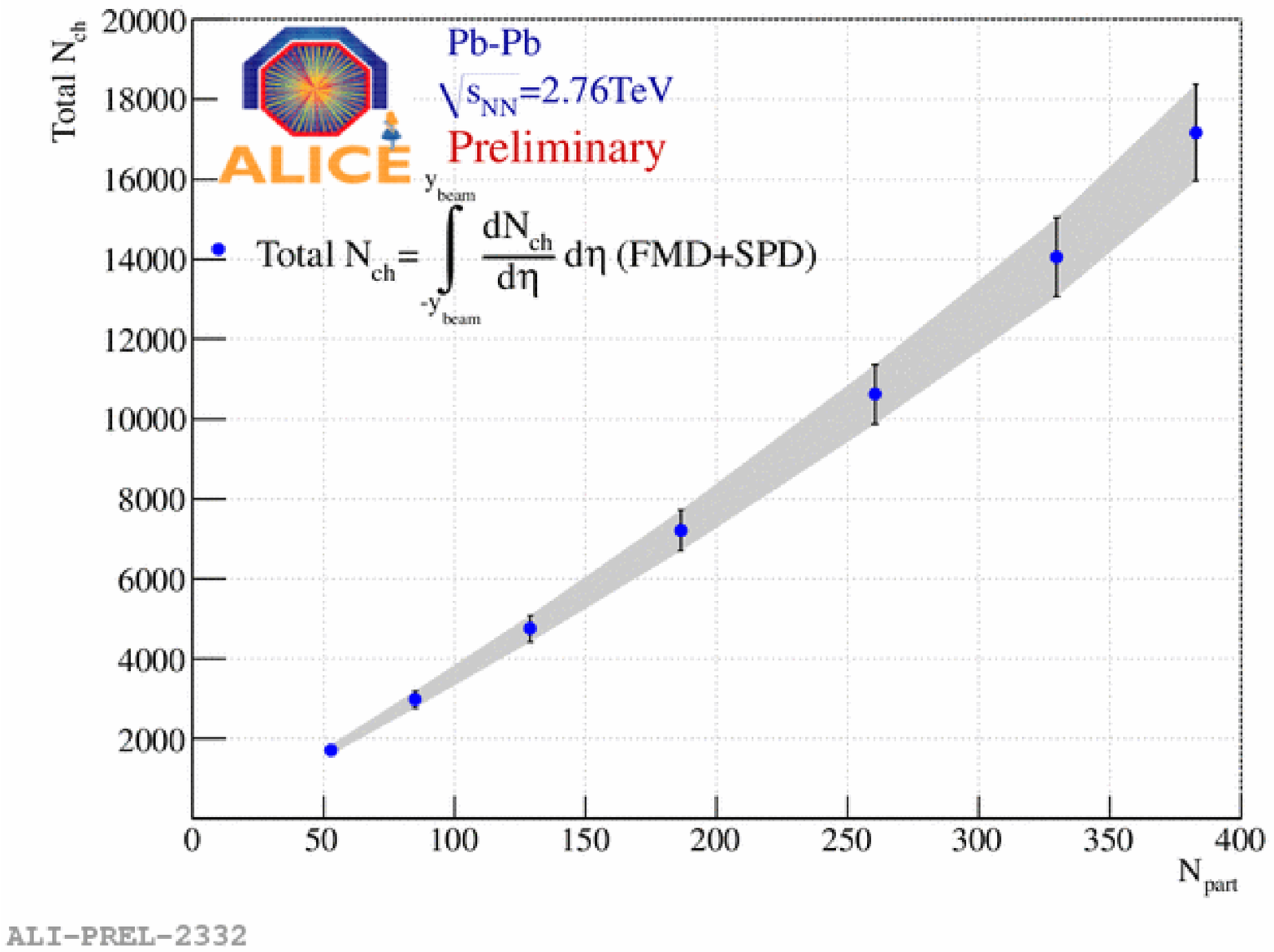}
  \caption{(left) $dN_{ch}/d\eta$ as a function of $\eta$ measured by the SPD (at mid-rapidity) FMD and VZERO detector (at forward-rapidity). (middle) The most central $dN_{ch}/d\eta$ as a function of $\eta-y_{beam}$. The data are fitted to a Gaussian and a linear fit. Also included in the figure is BRAHMS data for Au-Au collisions at 200~GeV. (right) Total number of charged particles N$_{ch}$ as a function of N$_{part}$. 
  \label{fig:fwdrap}}
\end{figure}

Observables at forward rapidity are interesting probes of the
properties of the initial state conditions, as for example the gluon
density or the Color Glass Condensate.  We measured the rapidity
dependence of the charged particle multiplicity, by extending the
rapidity coverage of the SPD measurement at mid-rapidity and
complementing it with two analyses at forward rapidity with the VZERO
and the FMD detector. The distributions are shown in
Fig.~\ref{fig:fwdrap} (left) for different centrality classes and the
the three measurements performed are in good agreement with each
other. The particle multiplicity integrated in a given rapidity range
increases with centrality with the same trend as the mid-rapidity
data. Yields at high rapidity are expected to be independent of the
energy, when viewed in the rest frame of one of the colliding
nuclei. So we compared our measurements with the ones at RHIC from the
BRAHMS experiment by shifting the rapidity by the beam rapidity (see
Fig.~\ref{fig:fwdrap} (middle)). We extend our measurement with two
extrapolations, a double Gaussian and a linear extrapolation. From the
comparison of the two energies it seems that the extended longitudinal
scaling may work even at LHC energies.  Thanks to those extrapolations
we have calculated the total charged particle multiplicity,
N$_{tot}$. N$_{tot}$ is proportional to the number of
participants N$_{part}$ (see Fig.~\ref{fig:fwdrap} (right)),
indicating that the pseudorapidity distributions get narrower for
more central collisions so that the width times the height of those
distributions is approximately constant with centrality, i.e. that the
increased particle production occurs mostly at mid rapidity.

\section{Interpretation, discussion and outlook} \label{sec:discuss}
This paper presents a summary of the first results on global event
properties measured by the ALICE Collaboration.  ALICE has measured
the charged particle pseudorapidity density at mid- and forward
rapidity and estimated the transverse energy.  The yields per
participant pairs show a consistent steady increase from peripheral to
central both for $dN_{ch}/d\eta$ and $dE_T/d\eta$. The centrality
dependence is strikingly similar to the one observed at RHIC energies.
Taking into account the measurements at lower energy, the $\sqrt
s_{NN}$ phenomenologically exhibits a power law scaling, stronger than
the logarithmic scaling proposed earlier both for $dN_{ch}/d\eta$ and
$dE_T/d\eta$.  The ratio $E_T/N_{ch}$ is flat with centrality, and its
increase in energy with respect to the top RHIC energy is consistent
with a 20\% increase in the mean transverse momentum.  The Bjorken
energy density is increased by about a factor 3
from the top RHIC energy.  In Fig.~\ref{fig:dNch} the data have been
compared to model calculations that describe RHIC measurements at
$\sqrt s_{NN}=200$~GeV and for which predictions at $\sqrt
s_{NN}=2.76$~TeV were available (see \cite{mul11,mul10} for a full
comparison).  The various models can be schematically divided in (i)
the perturbative-QCD-inspired Monte Carlo, based on HIJING, tuned to
7~TeV pp data. This type of models typically have a soft component
proportional to N$_{part}$ and a hard component proportional to
N$_{coll}$, partly motivating the parametrization used for the
ancestors (see Fig.~\ref{fig:vzeroglau}). Those models also include a
strong impact parameter dependence of parton shadowing, the one for
quarks fixed by the experimental data on DIS, the one from gluons
determined by the centrality dependence in heavy-ion
collisions. Others, (ii) so-called saturation model, also rely on pQCD
and use an initial-state gluon density to predict some
energy-dependent scale where quark and gluon density saturates
therefore limiting the number of produced partons, and in turn, of
particles. This results in a factorization of the energy and the
centrality dependence of the multiplicity, observed in the
experimental data.  In general theoretical models need some sort of
moderation mechanism to describe the centrality and energy evolution
of the multiplicity. Further constraints can be imposed on models when
describing also the forward rapidity region as well as the transverse
energy together.

\maketitle
\end{document}